\documentclass[conference, twocolumn]{IEEEtran}
\normalsize
\ifCLASSINFOpdf
\else
\fi

\usepackage{amsthm}
\usepackage{amsmath}
\usepackage{amssymb}
\usepackage{graphicx}
\usepackage{esint}
\usepackage{amsfonts}
\usepackage{cite}
\usepackage{balance}
\usepackage{caption}
\usepackage{subcaption}
\usepackage{epstopdf}
\usepackage{color}
\makeatletter
\usepackage{amssymb}
\usepackage{gensymb}
\usepackage{lipsum}
\usepackage{algorithm}
\usepackage{algpseudocode}
\usepackage[compatibility=false]{caption}
\usepackage[bottom]{footmisc}

\theoremstyle{plain}

\makeatother

\providecommand{\theoremname}{Theorem}

\theoremstyle{plain}

\makeatother

\providecommand{\lemmaname}{Lemma}

\begin{document}
\title{Relay-Based Blockage and Antenna Misalignment Mitigation in THz Wireless Communications}

\author{

Giorgos Stratidakis, Evangelos N. Papasotiriou, Haralampos Konstantinis, \\ Alexandros--Apostolos A. Boulogeorgos, and~Angeliki~Alexiou
\\

\begin{normalsize} 

Department of Digital Systems, University of Piraeus, Piraeus 18534, Greece.

\end{normalsize}

\\

\begin{normalsize} 

e-mail: (giostrat, vangpapasot, alexiou)@unipi.gr, xkonstantinis@gmail.com, al.boulogeorgos@ieee.org.

\end{normalsize} 

}

\maketitle	

\begin{abstract}
The proliferation of wireless devices in the recent years has caused spectrum shortage, which led the scientific community to explore the potential of using terahertz (THz) communications. However, THz systems suffer from severe path attenuation, blockage and antenna misalignment. In this paper, we present a relay-based blockage and antenna misalignment mitigation approach.
In more detail, two relay selection policies are employed, namely best and random relay selection. The system performance under both policies is evaluated and compared in terms of average throughput and the probability that throughput of a link is below a quality of service (QoS) threshold, using Monte~Carlo~simulations. It was observed that the effect of both blockage and misalignment can be mitigated using relays. Moreover, the gain of using relaying to mitigate blockage is much more significant.
\end{abstract}


\section{Introduction}\label{S:Intro}

To deal with the exponential increase of wireless devices and the exponentially increasing number of bandwidth demanding internet services the fifth generation (5G) wireless systems employ technological advances, such as full duplexing, massive multiple-input-multiple-output, as well as higher order modulations~\cite{J:THz_Enab_Wir_Sys_Bey_5G_Ult_Fast,A:LC_CR_vs_SS,WP:Wireless_Thz_system_architecture_for_networks_beyond_5G,PhD:Boulogeorgos}. Despite the paramount spectral efficiency increase, there is still a lack of flexibility in handling the tremendous amount of quality of service and experience-oriented data~\cite{Mag:Analytical_Performance_Assessment_of_THz_Wireless_Systems,Mag:THz_Techn_Del_Opt_Net_Qual,our_spawc_paper_2018,C:UserAssociationInUltraDenseTHzNetworks}. In this sense, the scientific community turned its attention to the THz frequency bands~\cite{Mag:Analytical_Performance_Assessment_of_THz_Wireless_Systems, Mag:THz_Techn_Del_Opt_Net_Qual,C:emeis_PIMRC_2019,C:THz_Enab_Wir_Sys_for_Bey_5G, C:ADistanceAndBWDependentAdaptiveModulationSchemeForTHzCommunications, our_VTC_paper,A:Error_analysis_of_mixed_THz_RF_wireless_systems,WP:Wireless_THz_system_applications_for_networks_beyond_5G,Stratidakis2019,Boulogeorgos2019}.

Despite the abundance of available spectrum in the THz band~\cite{Capacity_and_throughput_analysis_of_nanoscale_machine}; wireless systems operating in this band suffer from severe channel attenuation. In more detail, the THz propagation is highly dependent on the atmospheric conditions, since the major absorber of electromagnetic wave energy in this band is the atmospheric water-vapor~\cite{our_VTC_paper}, which causes discrete and deterministic loss to the transmitted signal. Hence, to improve the THz wireless system's performance high-gain transceiver antennas realizing extremely narrow beams (pencil-beams) are needed~\cite{Mag:Analytical_Performance_Assessment_of_THz_Wireless_Systems}. Furthermore, the effects of blockage and antenna misalignment in millimeter wave (mmWave) and terahertz (THz) systems, has been studied in several works. In more detail, the performance degradation due to antenna misalignment was evaluated in~\cite{Mag:Analytical_Performance_Assessment_of_THz_Wireless_Systems, paper_toyrkoi, The_impact_of_antenna_directivties,  C:THz_band_communications_Applications_research_challenges_and_standardization_activities,E.N.Papasotiriou2019,papasotiriou2019ergodic,A:Three_dimentional_end_to_end_modeling_and_analysis_for_graphene_enabled_THz_band_communications} (and references therein). In particular, in~\cite{paper_toyrkoi}~and~\cite{A:Three_dimentional_end_to_end_modeling_and_analysis_for_graphene_enabled_THz_band_communications}, the authors employed deterministic channel models  to evaluate the impact of the transceivers antenna misalignment, whereas, in~\cite{The_impact_of_antenna_directivties}, it was modeled as a part of the shadowing effect. On the other hand, in~\cite{Mag:Analytical_Performance_Assessment_of_THz_Wireless_Systems}~and~\cite{C:emeis_PIMRC_2019}, the effect of dynamic wind loads, weak earthquakes and thermal expansion, which lead to antenna misalignment was studied for THz wireless fiber extenders. Additionally, the small wavelength of millimter-wave (mmWave) and THz bands reduces the capability of signals to diffract around obstacles and is measured to attenuate the transmission path even by 40 dB~\cite{C:MMwave_Hum_Bl_73_GHz_with_a_Sim_Doub_Kn_Ed_Diffr_Mod_and_Ext_for_Dir_Ant, J:Diffr_in_mm_Sub_mm_Ind_Prop_Chan}, which combined with the THz transceivers pencil-beams could lead to line-of-sight (LoS) links interruption. This phenomenon is known as blockage. The effect of blockage in mmWave and THz wireless networks is commonly modeled, by employing the stochastic geometry approach. In more detail, the human/object blockers geometrical representation in two or three dimensional (2D/3D) scenarios is accomplished using the Poisson point process (PPP)~\cite{J:An_approx_for_intf_and_SIR_dens_in_thz_sys_w_atm_abs_dir_ant_and_bl,J:Int_and_SINR_in_mmwave_and_Thz_Com_Sys_with_Bl_and_Dir_Ant,
J:On_the_Temp_Eff_of_Mob_Bl_in_Urban_mmwave_Cellular_Scen}. Additionally, in~\cite{Let:Link_bl_an_for_ind_60ghz_radio_sys}~and~\cite{C:A_Hex_Grid_Based_Hum_Bl_Mod_for_the_5G_Low_THz}, the blocker's location is modeled through a uniform distribution. 
Finally, the effect of human blockage can be mitigated using relays to restore the link between the transmitter (TX) and the receiver (RX)~\cite{J:Bl_Rob_and_Eff_Sched_for_Dir_mmWave_WPANs,C:Rel_Sel_and_Sched_for_mmwave_Backhaul_in_Urban_Env,C:Deep_Learn_Based_Rel_Selec_In_D2D_mmWave_Com}.

To the best of the authors knowledge, there is no published contribution that reports a mitigation approach of the joint impact of antenna misalignment and blockage. Motivated by this, in this work, first we quantify the joint impact of antenna misalignment and human blockage in a wireless THz network, in which the user equipments (UEs) and blockers positions are generated by a PPP, while the antenna misalignment is modelled as a Rayleigh process, which is a common assumption in the highly directive THz wireless communications~\cite{Mag:Analytical_Performance_Assessment_of_THz_Wireless_Systems}. Then, in order to mitigate their joint impact, the use of relaying strategies is examined. In this direction, two relay selection strategies are discussed, namely the random and best relay selection. The performance of both strategies is assessed  through Monte Carlo simulations, employing the metrics of average throughput 
and cumulative distribution function (CDF) of the end-to-end capacity.

\section{System model}\label{sec:SM}

A wideband THz wireless network consisting of $N$ user equipments (UEs) is considered. It is assumed that there is no interference between the UEs and the transmission probability of each UE is $P_E$. Additionally, the UEs are equipped with highly directional antennas and the transceivers operate in half-duplex mode. The current topology describes a 2D network scenario, where the users are modelled as circles with radius $r_B$. Furthermore, their centers represent the UEs , whose locations are generated by a PPP with intensity $\lambda$. In order to avoid cases where two users are generated on top of each other, the minimum distance between them is set to~$2r_B$.

If a user interrupts the LoS connection of a link, it is considered as a blocker, which in turn reduces the signal-to-noise-ratio (SNR) of the link to zero. On the other hand, if the beams of the TX and the RX are not perfectly aligned, the received power and thus the SNR are reduced. 
To mitigate the joint impact of blockage and antenna misalignment, another UE, whose links with the TX and RX are not blocked or severely misaligned and satisfy the QoS throughput threshold, $C_m$ of the network, is set as a decode and forward relay (R) for this connection. Furthermore, the case with two maximum transmission hops is considered.

\begin{figure}
\centering\includegraphics[width=0.64\columnwidth,trim=0 0 0 0,clip=false]{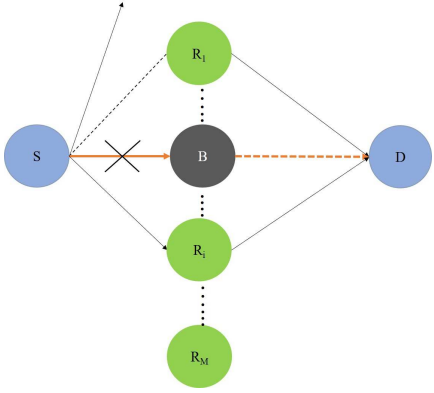}
\caption{Relaying for blockage and misalignment mitigation}
\label{fig:Example_of_Relaying}
\end{figure}

In Fig.~\ref{fig:Example_of_Relaying}, an intuitive example of the dual-hop relaying system topology is presented. As illustrated, one UE, namely the source $(S)$ transmits to the destination $(D)$, through an intermediate ($R$), because the direct link is obstructed by the blocker $(B)$. The $R$, is selected through a set ${\mathcal{R}=\{R_i|i=1,...,M\}}$ available $R$s.

The baseband equivalent received signal can be expressed~as
\begin{align}
y=h \ast x+ n,
\label{Eq:RX_signal}
\end{align}
where $x$, $h$ and $n$ are the complex transmitted signal, channel coefficient and additive white Gaussian noise (AWGN), respectively, for the three types of links, namely $(S,R_i)$, $(R_i,D)$ and $(S,D)$. Also, $\ast$ stands for the convolution operator. The channel coefficient can be expressed~as
\begin{align}
h = \begin{cases}
 \beta h_{l} h_{\phi},  & \beta=1  \\
 0,  & \beta=0 
\end{cases},
\label{Eq:h}
\end{align}
where $\beta=0$ and $\beta=1$ denote a blocked and an unblocked link, respectively. Moreover, $h_{l}$ stands for the deterministic path gain and $h_{\phi}$ is the antenna misalignment coefficient. 

From~\cite{EuCAP2018_cr_ver7}, $h_{l}$ can be obtained~as
\begin{align}
h_{l}(f,d)=\left(\frac{c}{4 \pi f d}\right) \sqrt{G_{t} G_{r}} \exp\left(-\frac{1}{2}k(f)d\right),
\label{Eq:Path_Gain}
\end{align}
where $f$ stands for frequency, $d$ is the transmission distance, $k(f)$ is the molecular absorption coefficient, while $G_{t}$ and $G_r$ stand for the TX and RX antenna gains,~respectively.

To accommodate the effect of misalignment fading $h_{\phi}$, the TX beam and the RX antenna effective area in the RX plane form circular discs. Furthermore, both discs are considered on the positive $x$-$y$ plane and the pointing error, $\rho=|\mathbf{\rho}|$ at the RX, expresses the radial distance of the transmission and reception beams. Also, without loss of generality, it assumed that $\rho$, is along the $x$-axis~\cite{Outage_capacity_FSO_2007}. Then, according to~\cite{Mag:Analytical_Performance_Assessment_of_THz_Wireless_Systems}~and~\cite{Outage_capacity_FSO_2007}, the antenna misalignment fading coefficient can be obtained~as
\begin{align}
h_{\phi}(\rho;d)=A_{o} \exp\left(- \frac{2 \rho^2}{R_{eq}^2} \right),
\label{Eq:Pointing_Error}
\end{align}
where $A_{o}$ is the fraction of the collected power by the RX at $\rho=0$, $R_{eq}$ is the equivalent TX beam radius at the RX obtained as in~\cite{Mag:Analytical_Performance_Assessment_of_THz_Wireless_Systems}. 
Also, by assuming independent identical Gaussian distributions for the horizontal and elevation displacements~\cite{Outage_capacity_FSO_2007},~\cite{arnon2003}, $\rho$ follows a Rayleigh distribution and its probability density function is expressed~as
\begin{align}
f_{\rho}(\rho) & =\frac{\rho}{\sigma_{s}^2} \exp\left(-\frac{\rho^2}{2 \sigma_{s}^2}\right), \text{ } \rho \geq 0,
\label{Eq:PDF_misalignment}
\end{align} 
where $\sigma_s$ is the jitter standard deviation.

\section{Relay Selection Strategies}\label{sec:RSA}
In this section two relay selection strategies are discussed.\\
\underline{Best relay selection strategy}: The selected relay of each blocked link is the UE that offers the highest throughput in the $S$-$R_i$ and $R_i$-$D$ links. This means that both the $S$-$R_i$ and $R_i$-$D$ links are not blocked and $C_{m}$ is satisfied.\\
\underline{Random relay selection strategy}: The $R_i$ of each $S$-$D$ blocked link is selected randomly with the discrete uniform distribution from among the UEs whose links with the $S$ and the $D$ are not blocked and satisfy $C_{m}$. 

\section{Performance Evaluation}

In order to quantify the performance and sustainability of the potential THz wireless links in the network, the average throughput and the probability $P_C$ that a link throughput $C$ is below the throughput threshold $C_{thr}$, are employed as metrics. For the THz wideband wireless communications the capacity of a link, can be obtained as~\cite{our_VTC_paper},~\cite{Capacity_and_throughput_analysis_of_nanoscale_machine},
\begin{align}
C=\int_0^B \log_2(1+\tilde{\gamma}(f)) \, df,
\label{Eq:Capacity}
\end{align}
where 
$\tilde{\gamma}(f)=\frac{P_t\left(\frac{c}{4 \pi f d}\right)^2 G_t G_r \exp\left(-k(f)d\right)}{N_o} h_{\phi}^2,$
where $P_t$ stands for the transmitted power spectral density (PSD), $N_o$ denotes the PSD of the AWGN and $B$ is the bandwidth. Moreover, the CDF of capacity, $P_C$ employing~\eqref{Eq:Capacity} can be obtained~as
\begin{align}
P_C= Pr(\tilde{C} \leq C_{thr}),
\label{Eq:PC}
\end{align}
where if the relayed link is used, $\tilde{C}=min(C_{S-R_i},C_{R_i-D})$, with $C_{S-R_i}$ and $C_{R_i-D}$ being the capacities of the $S$-$R_i$ and $R_i$-$D$ links and if the link is direct, $\tilde{C}= C_{S-D}$.

\section{Simulation results \& discussions}

In this section, Monte Carlo simulations of the average throughput and $P_C$ are delivered for different values of network UE densities and levels of antenna misalignment. 
In what follows, standard atmospheric conditions are assumed. Moreover, the transmission bandwidth, the central frequency, $\theta_{3dB}$, the network radius $R_N$, $C_{m}$, $r_B$ and $P_E$ are respectively set to $50$ GHz, $300$ GHz, $0.9^o$, $5$ m, ${1 \text{ Gb/s}}$, $0.2$ m and $60\%$. Furthermore, ${\lambda\in[0.3, 1.5]}$ $UEs/m^2$ and ${\sigma_s\in\{0, 0.05, 0.2\}}$ m. Also, note that $\sigma_s= 0$ m is the case without antenna misalignment. Finally, the parameter ${g=P_t G_t G_r/N_o=100\text{ dB}}$ is assumed, where $P_t$, $G_t$ and $G_r$ stand for the transmitted PSD, the gain of the TX, the gain of the RX and the PSD of the AWGN.

\begin{figure}
\centering\includegraphics[width=0.51\linewidth,trim=0 0 0 0,clip=false]{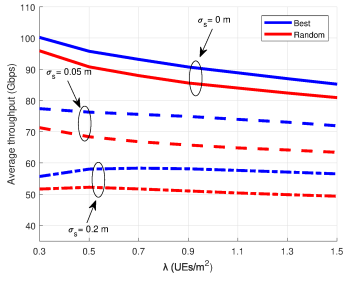}
\caption{Average throughput vs $\lambda$ for different values of $\sigma_s$.}
\label{fig:avg_thr}
\end{figure}

In Fig. \ref{fig:avg_thr}, the average throughput is depicted as a function of $\lambda$ for different values of $\sigma_s$. It observed that, for a given $\lambda$, as $\sigma_s$ increases, the average throughput decreases. Similarly, for a fixed $\sigma_s$, as $\lambda$ increases, the average throughput degrades. For example, as $\lambda$ increases from $0.3$ to $1.5$ $UEs/m^2$, the average throughput for both strategies with $\sigma_s= 0$ and $0.05$ m, decreases by $14.98$ and $5.49$ Gb/s, respectively. On the other hand, for $\sigma_s= 0.2$ m, as $\lambda$ increases from $0.3$ to $0.7$ $UEs/m^2$ the average throughput increases by $2.66$ Gb/s, while it decreases by $1.82$ Gb/s as $\lambda$ increases from $0.7$ to $1.5$ $UEs/m^2$. For $\lambda$ increasing from $0.3$ to $0.7$ $UEs/m^2$, a throughput increase is observed, as more UEs function as relays than as blockers, whereas for $\lambda$ increasing from $0.7$ to $1.5$ $UEs/m^2$, the throughput is reduced, as more UEs function as blockers than as relays. Furthermore, as $\sigma_s$ increases from $0$ to $0.05$ m and $0.05$ to $0.2$ m, the average throughput with $\lambda= 0.3$ $UEs/m^2$ decreases by $22.81$ and $21.71$ Gb/s, respectively. Furthermore, it is observed that since all links experience misalignment, its effect is harder to mitigate using relays than the effect of blockage.
Moreover, the difference between best relay and the random relay strategies increases when changing $\sigma_s$ from $0$ to $0.05$ m, while it decreases when changing $\sigma_s$ from $0.05$ to $0.2$ m. For example, for $\lambda= 0.3$ $UEs/m^2$ and $\sigma_s= 0$, $0.05$ and $0.2$ m the average throughput difference between the strategies is $5.13$, $9.16$ and ${7.1 \text{ Gb/s}}$, respectively. The reduction in the difference between the two strategies for high values $\sigma_s$ is caused by the reduction in the number of potential relays, due to severe misalignment. Thus, both strategies have to choose from a smaller pool of potential relays.

\begin{figure}
\centering
\begin{subfigure}[b]{0.51\linewidth}
   \includegraphics[width=1\linewidth]{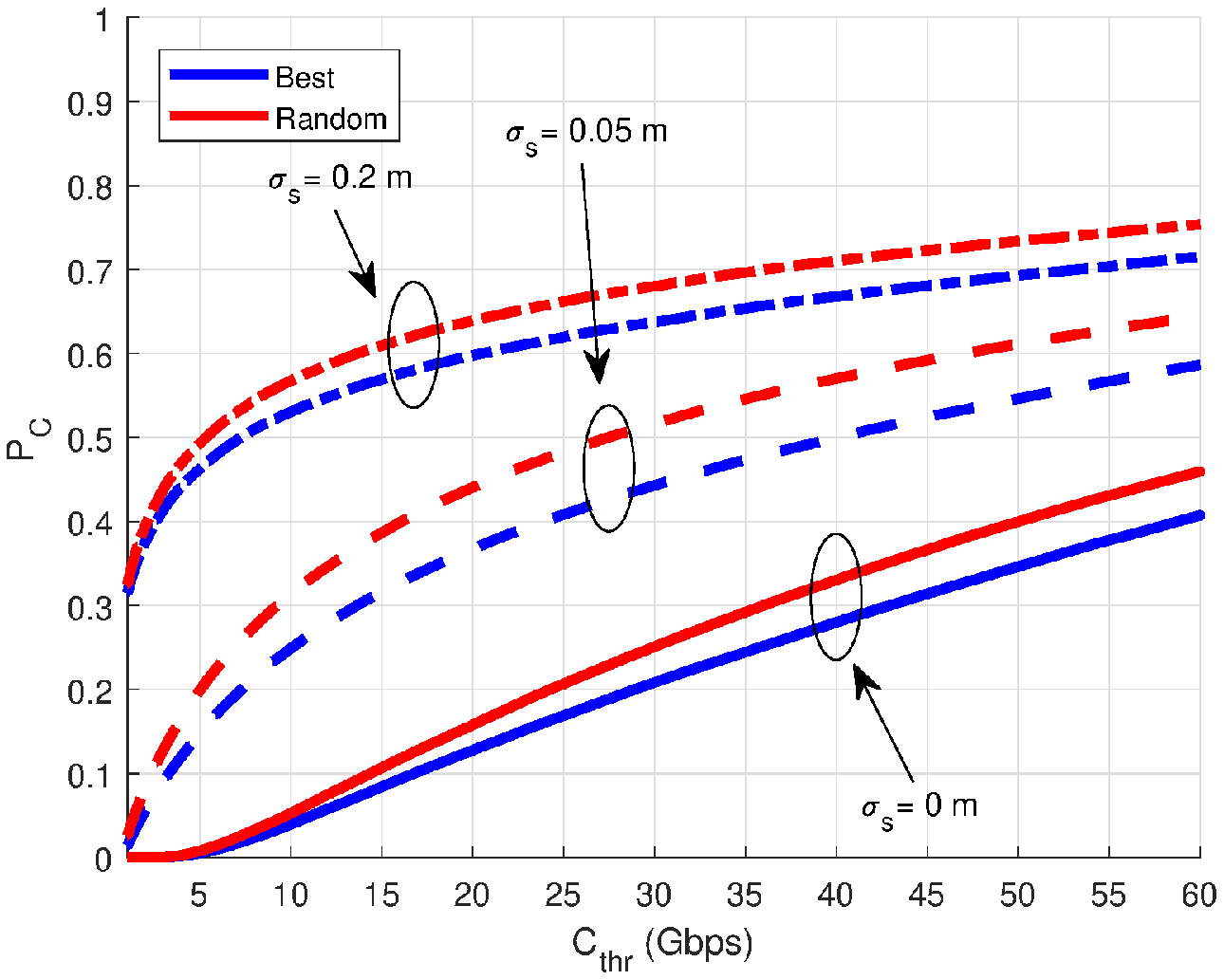}
   \caption{}
   \label{fig:OP_l_0.3}
\end{subfigure}

\begin{subfigure}[b]{0.51\linewidth}
   \includegraphics[width=1\linewidth]{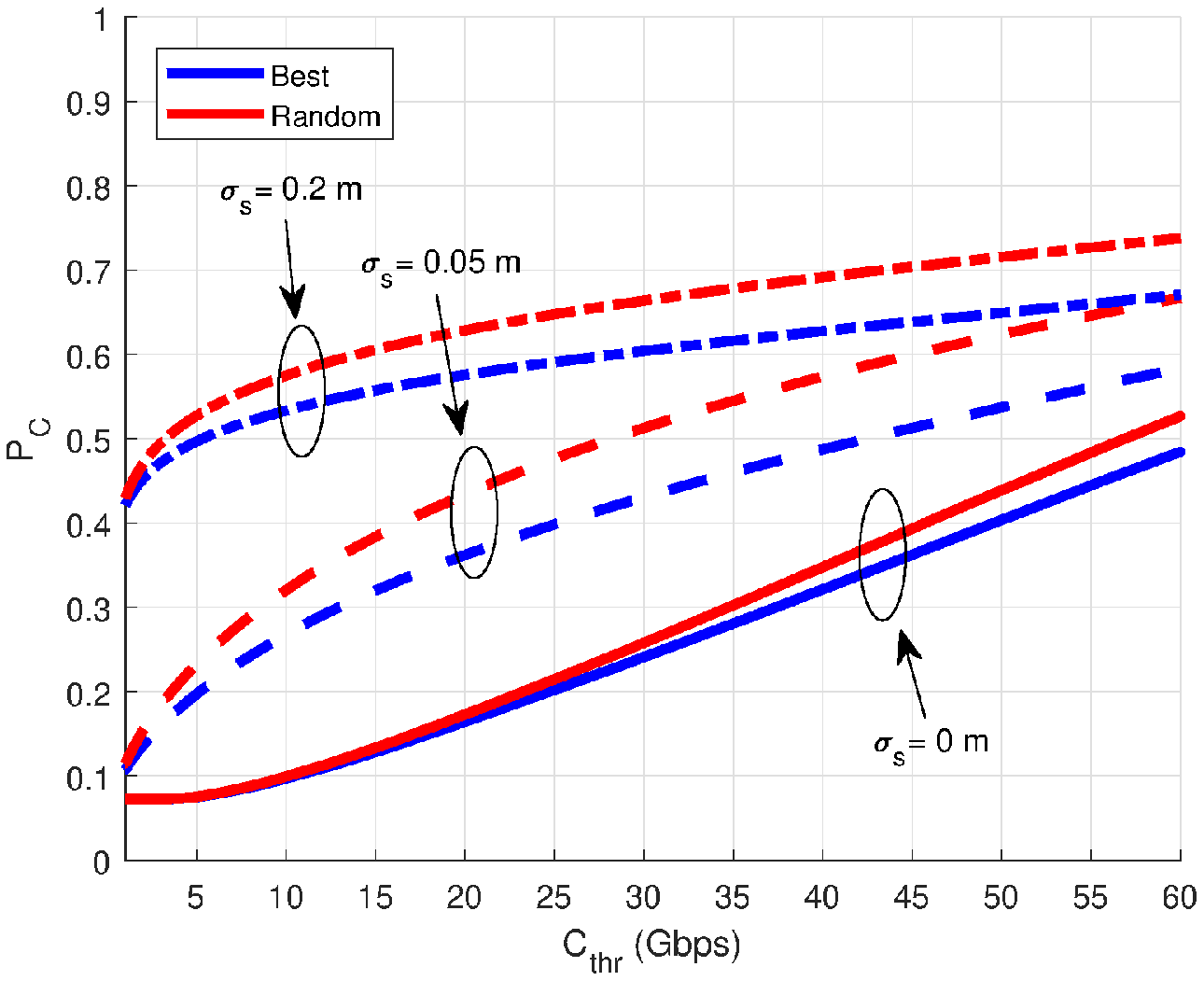}
   \caption{}
   \label{fig:OP_l_1.5}
\end{subfigure}
\caption[$P_C$ vs target throughput for different levels of $\sigma_s$]{$P_C$ vs target throughput for different levels of $\sigma_s$: (a) $\lambda=0.3$ $UEs/m^2$ and (b) $\lambda=1.5$ $UEs/m^2$}
\label{fig:OP_vs_tgt_thr}
\end{figure}

Fig.~\ref{fig:OP_vs_tgt_thr} shows $P_C$ as a function of the target throughput for different values of $\lambda$ and $\sigma_s$. As expected, for a fixed $\lambda$, as $\sigma_s$ increases, $P_C$ also increases. Moreover, for a fixed $\lambda$, as $\sigma_s$ increases, $P_C$ also increases. Furthermore, as expected the best relaying achieves the lowest $P_C$ for each combination of $\lambda$ and $\sigma_s$. For example, for $\lambda=0.3$ $UEs/m^2$, ${\sigma_s=0.05\text{ m}}$ and target throughputs of $30$ and ${45\text{ Gb/s}}$ the $P_C$ difference between the two strategies is $0.071$ and $0.069$, respectively, whereas for the same parameters and ${\sigma_s=0.2\text{ m}}$ their $P_C$ difference is $0.0432$ and $0.0418$, respectively. Moreover, for $\lambda=1.5$ $UEs/m^2$, ${\sigma_s=0.05\text{ m}}$, and target throughputs of $30$ and ${45\text{ Gb/s}}$, the $P_C$ difference between the two strategies is $ 0.082$ and $ 0.087$, respectively, whereas for the same parameters and ${\sigma_s=0.2\text{ m}}$ their $P_C$ difference is $ 0.06$ and $ 0.066$, respectively. Finally, it is observed that the best relay performance gain in respect to the random relay one, decreases as $\sigma_s$ increases.

\section{Conclusions}
In this work, the joint impact of blockage and antenna misalignment in THz wireless networks capable of using two-hop relaying was quantified. In this direction, the best and random relay selection strategies were reported. It was observed that the increase of the PPP intensity $\lambda$ increases the number of both blockers and relays, while the increase of the jitter standard deviation $\sigma_s$ decreases the number of UEs that can be used as relays. Our results indicate that, in general, the  best relay selection strategy outperforms the random one. Furthermore, using relaying is much more efficient in mitigating the effect of blockage than misalignment. Finally, the importance of taking into account both the antenna misalignment and blockage, when evaluating the performance of THz wireless networks, was highlighted.

\section*{Acknowledgment}
This work has received funding from the European Commission's Horizon 2020 research and innovation programme TERRANOVA under grant agreement No. $761794$.

\balance
\bibliographystyle{IEEEtran}
\bibliography{IEEEabrv,References}
\end{document}